\documentclass[aps,prl,twocolumn,nofootinbib]{revtex4}
\usepackage{graphicx}
\usepackage{bm}
\def\be{\begin{equation}}
\def\ee{\end{equation}}
\def\ed{\end{document}}
\def\mpl{M_{\rm Pl}}
\def\mm{\widetilde{m}_\phi^2}

\begin{document}

\title{Time Variation of the Fine Structure Constant Driven by Quintessence}
\author{Luis Anchordoqui$^a$ and Haim Goldberg$^{a,b}$}
\affiliation{$^a$ Department of Physics, Northeastern University, Boston,
MA 02115\\
$^b$ Center for Theoretical Physics, Massachusetts Institute
of Technology, Cambridge, MA 02139
}

\begin{abstract}
There are indications from the study of quasar absorption spectra that the fine structure
constant $\alpha$ may have been measurably smaller for redshifts $z>2.$ Analyses of other data
$(^{149}$Sm fission rate for the Oklo natural reactor, variation of $^{187}$Re $\beta$-decay rate
in meteorite studies, atomic clock measurements) which probe variations of $\alpha$ in the more
recent past imply much smaller deviations from its present value. In this work we tie the variation
of $\alpha$ to the evolution of the quintessence field proposed by Albrecht and Skordis, and show
that agreement with all these data, as well as consistency with WMAP observations, can be achieved
for a range of parameters. Some definite predictions follow for upcoming space missions
searching for violations of the equivalence principle.\\

\centerline{NUB-3237-TH-03~~~~~~MIT-CTP-3385~~~~~~hep-ph/0306084}
\end{abstract}

\maketitle

\section{Introduction}
Independent observations of a number of absorption systems
in the spectra of distant quasars (QSOs) seem to indicate that $\alpha$,
the fine structure constant of quantum electrodynamics, is slowly
increasing over cosmological time scales~\cite{Webb:1998cq,Bekenstein:2003zb}.
Specifically, the experiments indicate that averaged over
redshifts $0.2<z<3.7,$ there is a 5.7$\sigma$ deviation of the
fine structure constant from its present value, namely
$\Delta\alpha/\alpha = -0.57\pm 0.10\times 10^{-5}$~\cite{alpha}. On the other
hand, terrestrial and solar system measurements provide several
constraints on recent rate of variation of $\alpha:$ (1) Analyses
of the resonant fission reaction rate in the naturally occurring
reactor at Oklo in Gabon provide a bound
$-0.9 \times 10^{-7}  < \Delta\alpha/\alpha < 1.2 \times 10^{-7}$
over the era $z<0.14$ at the 95\% CL~\cite{Damour:1996zw}.
(2) Based on plausible assumptions, new estimates of the age of
iron meteorites $(z\approx 0.45)$ combined with a measurement of
the Os/Re ratio resulting from the radioactive decay $^{187}$Re
$\rightarrow$ $^{187}$Os have allowed a narrowing in the
uncertainty of the average decay rate over the age of the
meteorite~\cite{Smoliar}. This has been translated~\cite{Olive:2002tz} into a strong
bound $\Delta\alpha/\alpha<3\times 10^{-7}$, following the
original suggestion of Peebles and Dicke~\cite{Peebles}. (3) Recently,
three years of observations of hyperfine spectra using atomic
fountain clocks have allowed a 1$\sigma$ bound $\dot\alpha/\alpha
< 1.6\times 10^{-15}$ yr$^{-1}$ for the present time-rate of
variation of $\alpha$~\cite{Marion:2002iw}. (4) Additional bounds in this
category have been derived~\cite{Sisterna:et}, but these are weaker than
the ones listed above. Finally, there are  constraints resulting
from cosmological considerations: (1) Limits on the
temperature fluctuations of the cosmic microwave background
(CMB) could lead to a measurement with experimental sensitivity of
$|\Delta\alpha/\alpha|< 10^{-2} - 10^{-3}$, at
$z \sim 1000$~\cite{Hannestad:1998xp}. Analysis of
data from
the Wilkinson Microwave Anisotropy Probe (WMAP) provides a bound
$-0.06<\Delta\alpha/\alpha<0.02$ at 95\% CL~\cite{Martins:2003pe}. (2) Big bang
nucleosynthesis (BBN) considerations place bounds on
 $|\Delta\alpha/\alpha|$ on the same order of magnitude as those from CMB,
though at much larger redshift, $z \sim 10^9 - 10^{10}$~\cite{Kolb:1985sj}.

Over the last years, a second set of observations (most recently from
WMAP~\cite{Spergel:2003cb}) has accumulated which indicate that the
universe is spatially flat to within 1\%. In addition, luminosity
distance measurements of  Type Ia supernovae strongly imply
the presence of some unknown form of energy density, related to
otherwise empty space, which appears to dominate the recent
gravitational dynamics of the universe and yields a stage of
cosmic acceleration~\cite{Riess:1998cb}. We still have no solid
clues as to the nature of such dark energy (or perhaps more
accurately dark pressure), but in recent years it has been associated
with a dynamical scalar field $\phi$ evolving in a potential $V(\phi)$~\cite{Ratra:1987rm},
generally called ``quintessence''~\cite{Zlatev:1998tr}.

The universality of gravitational interactions implies that one
may expect the Lagrangian below the Planck scale to contain
non-renomalizable couplings of $\phi$ to standard model fields~\cite{Carroll:1998zi}.
In particular, the free Lagrangian for the electromagnetic field tensor
$F_{\mu\nu}$ will be modified to
\be
 \widetilde{\cal L}_{\rm em} = -\frac{1}{4}\ Z_F(\phi/\mpl)\ F_{\mu\nu}F^{\mu\nu}\ \ ,
\ee
with $\mpl = (8\pi G)^{-1/2},$ the reduced Planck mass.
On expansion about the present value $\phi_0$ of $\phi$, this becomes
\be
\widetilde{\cal L}_{\rm em} = -\frac{1}{4}\ (1+ \kappa\frac{\Delta\phi}{\mpl}\ + \ldots) \,
F_{\mu\nu}F^{\mu\nu}\ \ ,
\label{expansion}
\ee with $\Delta\phi = \phi-\phi_0$
and $\kappa\equiv \left. \partial_\phi Z_F \right|_{\phi_0}.$ The field
renormalization $A_{\mu}\rightarrow A_{\mu}/Z_F^{1/2}$ to obtain
a canonical kinetic energy, generates  an effective charge
$e/Z_F^{1/2}.$ Expansion to linear order about the present value
$e_0$, leads to
\be \frac{\Delta \alpha}{\alpha} = - \kappa
\frac{\Delta\phi}{\mpl}\ \ .
\label{alpha}
\ee
Compatibility between the Oklo/metorite/atomic clock and QSO
measurements can greatly  constrain  the dynamics of
$\phi$: its evolution should slow considerably between the quasar
era and the present epoch~\cite{Chiba:2001er}.

If the field $\phi$ driving the variation in $\alpha$ is a
quintessence field, then its evolution is further constrained by
observation. In particular, it must  provide about 70\% of the
total energy density at present. Its equation of state, $w_\phi
\equiv p_\phi/\rho_\phi$ ($p_\phi\equiv$ pressure,
$\rho_\phi\equiv$ energy density), is most strongly constrained
by WMAP observations: $w_\phi < - 0.78$ at the
95\%CL~\cite{Spergel:2003cb}. Additionally, radiation dominance at
the time of BBN must be maintained -- no more than 20\% of the
energy density at that time can reside in
quintessence~\cite{Olive:1999ij}.

In this work we search for a model of quintessence in which the
evolution of the scalar field $\phi$ in its potential $V(\phi)$
drives the variation in $\alpha$~\cite{Kostelecky:2002ca}. The particular case where
$\phi$ is the dilaton was examined in ~\cite{Damour:2002mi}. With
such a dynamics, however, the universal coupling of the dilaton
greatly constrains the variation in $\alpha,$ making it difficult
to comply with the QSO data. An alternative
proposal~\cite{Wetterich:2003jt}, where the dynamics of the
scalar field resides in a non-trivial K\"ahler potential, can
provide a variation of $\alpha$ compatible with observation. However,
in this work we wish to pursue a path based on sums of exponential
potentials, which can be more explicitly linked to string theories (more on this below).
It is also important to stress that our approach  differs fundamentally
from models where the variations of the scalar field are
primarily driven by its couplings to non-relativistic
matter~\cite{Olive:2001vz,Gardner:2003nw} (generically referred
to as Bekenstein-type models~\cite{Bekenstein:eu}).

Before proceeding, we take note of a discussion~\cite{Banks:2001qc}
which relates this type of
variation in $\alpha$ to a large shift in the cosmological
constant, and hence questions its viability.
In the spirit of~\cite{Langacker:2001td} we adopt here a more
wait-and-see position, since present field-theoretic
considerations all require fine-tuning to solve the cosmological
constant problem, and  may need to be totally supplanted (perhaps
by self-tuning mechanisms, such as described
in~\cite{Arkani-Hamed:2000eg}).

\section{Albrecht-Skordis Cosmology}

\subsection{Quintessence Phenomenology}

An interesting model for quintessence has been presented by
Albrecht and Skordis~\cite{Albrecht:1999rm}. The scalar field evolves in a
potential (hereafter we adopt natural units $8 \pi G = 1$)
\be
V(\phi) = V_p(\phi)\,\,\exp[-\lambda\, \phi]\,\,,
\label{vphi}
\ee
with an economic polynomial factor,
\be
V_p (\phi) = (\phi - B)^{\beta} + A \,,
\ee
in which the constants $A, B, \beta$ and $\lambda$ are phenomenologically
determined. Because of the polynomial factor, this potential
differs in a critical manner from the
much-studied pure exponential: although the tracking properties are similar,
it allows sufficient radiation dominance during BBN while evolving to
quintessence dominance in the present epoch, all largely independent
of initial conditions. For the particular case $\beta=2,$
the field is trapped in a minimum yielding a permanently accelerated universe
if $\lambda^2 \, A <1,$  whereas if $\lambda^2 \,A>1$ (and for a small
region $0 <1- \lambda^2 \, A \ll 1$),
the accelerated era is transient~\cite{Barrow:2000nc}. In what follows we
set $\beta =2.$

Exponential potentials are ubiquitous in 4-D field theory
descendants of string/M theory~\cite{Gasperini:2001pc}.
Additionally, sums of exponentials have been
proposed~\cite{deCarlos:1992da} in earlier attempts at
stabilizing the dilaton and allowing supersymmetry breaking
through gaugino condensates. In order to avoid the difficulties
associated with identifying quintessence as the
dilaton~\cite{Damour:2002mi}, in what follows  we associate
$\phi$ with  moduli related to compactification, since, unlike
dilatons, the latter need not be universally coupled to matter
and gauge fields. The origin of the polynomial factor form of the
potential in Eq.~(\ref{vphi}) can be linked to a non-trivial
K\"ahler term in an effective 4-D supergravity
theory~\cite{Copeland:2000vh}. However, in such a case an
explicit connection to string theory has not been successfully
established. On the other hand, string/M theory does provide
motivation for superpotentials which are sums of
exponentials~\cite{Cvetic:1999xp}. A
much-discussed~\cite{Bremer:1998zp}  recent example occurs in 11-dimensional
supergravity, with a geometry consisting of a warped product of
our 4 dimensional space-time and  an internal  compact
7-dimensional  hyperbolic manifold whose volume is proportional to
the dual of the field strength. Upon dimensional reduction, the
effective potential in 4 dimensions consists of two exponentials,
the first proportional to the 7-dimensional Ricci scalar, the
second to the volume of the compact space. The exponents are
proportional to the breathing modulus of the warp factor.

With this in mind, we can think of the Albrecht-Skordis potential
as the limiting case of three exponentials
\begin{eqnarray}
V(\phi) & = &   C\,e^{-(\lambda+\epsilon)(\phi-B)} - 2C\,(1-\epsilon^2 A/2)\,e^{-\lambda(\phi-B)}
\nonumber \\
& + & C\,e^{-(\lambda-\epsilon)(\phi-B)}\,,
\end{eqnarray}
where $ \epsilon^2 C = e^{-\lambda B},$ and  $\epsilon^2\ll 1.$
In  order that the reduction to a 4-D theory generate three
exponentials with a single modulus, it is necessary to split  the
compactification of the internal space so that the resulting internal manifolds
have different warping factors with a single collective coordinate.
Additional dilatonic degrees of freedom can arise if one starts from 10-D string theory
(such as Type IIA or Type IIB), and these need to be constrained or stabilized.
We are currently studying these possibilities and the results will be
presented elsewhere~\cite{inprogress}.

\subsection{Cosmological Evolution of $\alpha$}

In order to follow the evolution of the fine structure constant, we need to trace the
temporal behavior of $\phi$ since the quasar epoch, while at the same time requiring
that the field provide a  successful model for quintessence.
This evolution follows from the 4-D effective action
\be
S = \int d^4x \,\sqrt{-g} \left\{ \frac{R}{2} + {\cal L}_\phi +
{\cal L}_{\rm vis} + [Z_F(\phi) - 1]\, {\cal L}_{\rm em} \right \}.
\ee
As usual, $R$ is the Ricci scalar derived
from the metric tensor $g_{\mu\nu}$,
\be
{\cal L}_\phi = \frac{1}{2} \partial_\mu \phi\,\, \partial^\mu \phi - V(\phi),
\ee
${\cal L}_{\rm vis}$ is the
Lagrangian of visible matter (baryons, photons, and also baryonic and neutrino dark matter), and
\be
{\cal L}_{\rm em} = - \frac{1}{4} \,F_{\mu\nu}\, F^{\mu\nu} \,.
\ee
Note that we have omitted terms coupling $\phi$ to the other matter fields because we are searching
for a non-Bekenstein solution, i.e., the potential is primarily driving the evolution of $\phi$.

The equation of motion for $\phi$ then reads
\be \ddot \phi + 3
\,H \,\dot \phi = - \frac{\partial V}{\partial\phi} \,,
\ee
where
\be
H^2 \equiv \left(\frac{\dot{a}}{a}\right)^2 = \frac{1}{3} \,
\left[ \frac{1}{2}\, \dot\phi^2 + V(\phi) + \rho_{\rm m} +
\rho_{\rm rad}\right],
\ee
with $\rho_{\rm m}$ and $\rho_{\rm r}$ the
matter and radiation energy densities, respectively. It is more convenient to
consider the evolution in \be x=\ln a = - \ln (1 +z) \,, \ee with
the present value of the scale parameter $a_0 = 1.$ Denoting by a prime
derivatives with respect to $x,$ the equation of motion for $\phi$ becomes
\be
 \frac{\phi^{\prime\prime}}{1 -\, \phi^{\prime^2}\!/6} \,\, +  3 \,\phi^\prime +
\frac{\rho^\prime \,\phi^\prime/2 \,\,+ \,3 \,\,\partial_\phi
V}{V+\rho} = 0\,,
\label{motion}
\ee where $\rho = \rho_{\rm m} + \rho_{\rm r}.$ Quantities of importance are the dark energy density
\be
\rho_\phi = \frac{1}{2}\, H^2 \,\phi^{\prime
2} + V \,, \ee
generally expressed in units of the critical density ($\Omega\equiv\rho/\rho_{\rm c}$)
\be
\Omega_\phi = \frac{\rho_\phi}{3 H^2}\,,
\ee
and
\be
H^2 = \left[\frac{V +
\rho}{3}\right]\,\left[1 -
\frac{\phi^{\prime 2}}{6}\right]^{-1}\,\,.
\ee
The equation of state is
\be
w_\phi = \left[\frac{H^2 \,\,\phi^{\prime 2}}{2} -
V\right]\,\left[\frac{H^2 \,\,\phi^{\prime 2}}{2}+
V\right]^{-1}\,.
\ee
For $\rho_{\rm m}$ and $\rho_{\rm r}$ we adopt the expressions
\be
\rho_{\rm m} = {\cal C}\, e^{-3x}, \,\,\,\,\, \rho_{\rm r} = 10^{-4}\, {\cal C}\, e^{-4x} \,
f(x) \,,
\ee
where ${\cal C}=\Omega_{m,0}\,\rho_{{\rm c},0},$
and $f(x)$ parameterizes the $x$-dependent number of radiation degrees of
freedom. In order to interpolate the various thresholds appearing prior to
recombination (among others, QCD and electroweak), we adopt a convenient
phenomenological form,
\be
f(x)=\exp(-x/15).
\ee
In natural units  $\rho_{{\rm c},0} = 2.3 \times 10^{-120} \,h_0^2.$ Finally, we
set our constants $\Omega_{{\rm m},0}=0.3,$ $h_0=0.7$ in accordance with WMAP
observations~\cite{Spergel:2003cb}.

\begin{figure}
\begin{center}
\includegraphics[height=8.5cm]{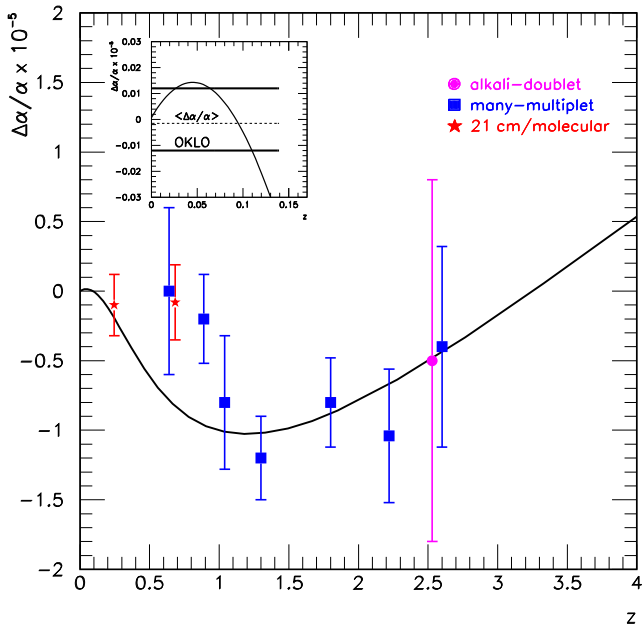}
\caption{The solid line indicates the variation of $\Delta\alpha/\alpha$ for $\mm=52.5$, $\lambda=8.5,$
$B=32.0,$ and $\kappa=-3.3\times 10^{-4}.$ The cosmological evolution of
$\alpha$ is superimposed over a binned-data
sample from 72 QSO absorption systems: The  points indicated by  $\star$ correspond to two HI 21 cm
and molecular absorption systems~\cite{Murphy:2001nu}. Those points assume no change on the proton
$g$-factor, and therefore should be interpreted with caution. The 7 squares are binned results for 49
QSO absorption systems~\cite{Murphy:2000pz}. The lower redshift points (below $z \approx 1.6$) are based
on (MgII/FeII) and the higher redhsift points on (ZnII, CrII, NiII, AlIII, AlII, SiII). The single point
indicated by $\bullet$ represents the average over 21 QSO SiIV absorption doublets using the alkali
doublet method~\cite{Murphy:2000ns}. The embedded box details the behavior of $\Delta\alpha/\alpha$
for small values of $z$. The solid horizontal  lines indicate the bound derived from the nuclear reactor
at Oklo, whereas the dotted line in the middle indicates
the predicted average value of $\Delta\alpha/\alpha$
within $0 < z < 0.14$.}
\label{fig1}
\end{center}
\end{figure}

As described in~\cite{Albrecht:1999rm}, for a wide range of parameters this
potential allows a plausible cosmological behavior
independent of initial conditions. For definiteness, in this work we take
the initial kinetic energy in the field $\phi$ to equal its initial potential energy.
The remaining degrees of freedom $\lambda$ and $A$ will be used in order to study the
variation of the fine structure constant over the history of the universe.
We have found that the recent variation of $\phi$ is most directly controlled by the
curvature of the potential at its minimum. This is given by the mass of the scalar field,
\be
m_{\phi}^2 = 2 \ \,(1-{\cal K}) \,e^{-\lambda B - {\cal K}}\ \ ,
\ee
where
\be
{\cal K}   = 1 - \sqrt{1-A\lambda^2} \ \ .
\ee
We will state our results in terms of
\be
\widetilde{m}_\phi^2 \equiv \frac{m_\phi^2}{H_0^2} = \frac{3\, \Omega_{\phi,0}\,\,
\lambda^2 \,\,(1-{\cal K})}{{\cal K}} \ \ .
\ee
For fixed values of $\widetilde{m}_\phi^2$ and $\lambda,$ the value of $B$ is fixed by
requiring that the current dark energy density constitute a fraction  $\Omega_{\phi,0}$
of the critical density.

Now, Eq.~(\ref{motion}) is integrated for a range of values of $\lambda$ and
$\mm$, from $a=10^{-30}$ to the present epoch. Of these, a small subset has been found to be of
interest with respect to the data on the  variation of $\alpha$ presented in the
Introduction. Within this subset, two types of solution can be identified: in the first,
the variation of $\alpha$ shows an oscillatory behavior between the present and QSO epochs,
which reflects (through Eq.~(\ref{alpha})) the oscillatory behavior of $\phi$ in the potential well;
in the second,
the motion of $\phi$ as it comes to its present value is overdamped,
so that the change in $\alpha$ is a monotonically
decreasing function of $z.$

In Fig.~\ref{fig1} we show an example of the oscillatory behavior, for the set of
parameters $\mm=52.5$, $\lambda=8.5,$
$B=32.0$, and  $\kappa=-3.3\times 10^{-4}.$ Although the fit to the QSO data is not particularly good
($\chi^2/{\rm d.o.f.}= 16/7$), a solution of this type (with some $V_p$ of higher order in
Eq.~(\ref{vphi})) can lead to an interesting prediction for future observations:
$|\Delta\alpha/\alpha|$ diminishes with increasing redshift beyond $z=3$. The fit is consistent with the
Oklo bound (see the inset), but it fails to comply with the new meteorite bound.

\begin{figure}
\begin{center}
\includegraphics[height=8.5cm]{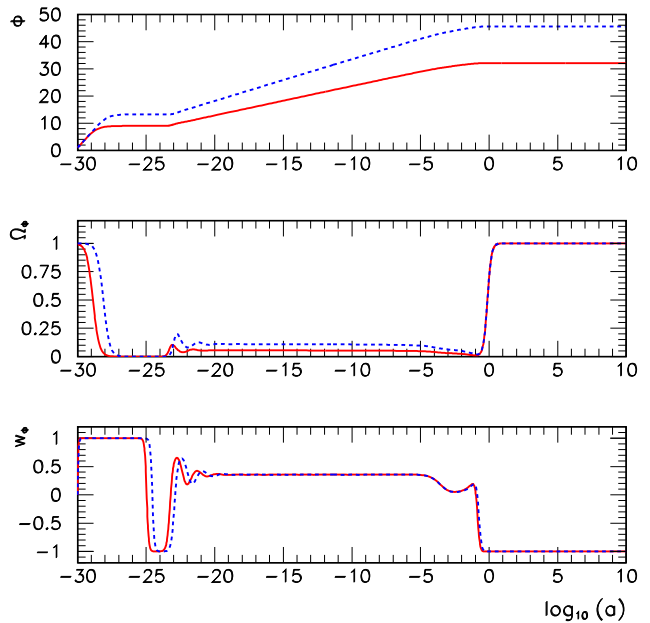}
\caption{The upper panel shows the evolution of $\phi$ for $\mm=52.5$, $\lambda=8.5,$
$B=32.0$ (solid line) and $\mm=10.5,$ $\lambda=6,$ $B=45.4$
(dotted line) as a function of ${\rm \log}_{10}(a)$. Today corresponds to $a=1$, for BBN
$a \approx 10^{-10}$, and for the Planck epoch $a \approx 10^{-30}$. The second panel shows the
evolution of $\Omega_\phi$ for the $V$-parameters described above. The lower panel
shows the evolution of the equation of state $w_\phi$ (same conventions than the upper panel).
These two solutions are consistent with the BBN requirement $\Omega_\phi < 0.2$~\cite{Olive:1999ij},
show the established radiation and matter dominated
epochs, and at the end yield an accelerated quintessence era.}
\label{fig2}
\end{center}
\end{figure}

Further comparison with data must include compliance with measured bounds on violation
of the equivalence principle, in the form of limits on composition dependent
inertial forces~\cite{Dvali:2001dd,Damour:2002mi,Olive:2001vz}. These can translate into an
upper bound on $\kappa.$ Along these lines, Olive and Pospelov~\cite{Olive:2001vz} parametrize the
$\phi$-dependent modification of the electromagnetic Lagrangian via an equation analogous to
Eq.~(\ref{expansion}),
\be
\widetilde{\cal L}_{\rm em} = -\frac{1}{4}\ (1+ \zeta_F\frac{\Delta\phi}{M_*}\ + \ldots) \,
F_{\mu\nu}F^{\mu\nu}\ \ ,
\label{op}
\ee
where $M_*$ is the analogue of $\mpl$ in the $\phi$ sector, and the field $\phi$ in Eq.~(\ref{op}) is
defined to have a canonical kinetic energy.
Comparing Eqs.~(\ref{expansion}) and (\ref{op}), we find $\kappa = \zeta_F/\sqrt{2\omega},$
where $\omega\equiv M_*^2/2\mpl^2.$ The limit derived in~\cite{Olive:2001vz}
$\zeta_F/\sqrt{\omega} < 10^{-3},$ requires
\be
\kappa < 7 \times 10^{-4}\ \ .
\label{kappa}
\ee
We can see that our value,  $\kappa = 3.3 \times 10^{-4}$, is consistent with the one given in
Eq.~(\ref{kappa}), but close: taking this model seriously would suggest that composition-dependent
inertial forces could be observed in an improved round of experiments.

The third constraint mentioned in the Introduction is the variation of atomic clock measurements.
In terms of our variables, the fractional change in the present epoch is
\be
\frac{\dot\alpha}{\alpha} = \kappa H_0 \phi^{\prime}_0\ \ ,
\ee
where for this set of parameters $\phi^{\prime}_0 = -7.8 \times 10^{-3}.$ This implies
$\dot\alpha/\alpha = 1.8\times 10^{-16}$ yr$^{-1}$, well within the bound stated in the Introduction.

Finally, as can be seen in Fig.~\ref{fig2}, the change in $\phi$ between BBN and the present is $\sim 10$.
On the assumption that the dominant variation in $Z_F(\phi)$ is linear over this domain, the fractional
change in $\alpha$ is well within the sensitivity of present analysis. A similar statement holds for
the CMB~\cite{Huey:2001ku}.

\begin{figure}
\begin{center}
\includegraphics[height=8.5cm]{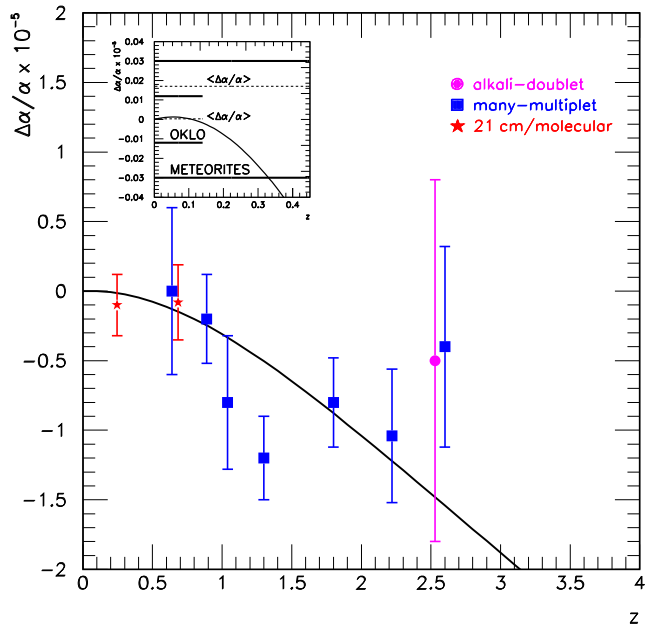}
\caption{Cosmological evolution of $\Delta\alpha/\alpha$ for $\mm=10.5,$ $\lambda=6,$
$B=45.4,$ and
$\kappa = 2.1 \times 10^{-4}.$ The experimental data points are those of Fig.~\ref{fig1}. The
embedded box displays both bounds from the Oklo reactor and from meteorite analyses
(horizontal solid lines). The predicted average value of $\Delta\alpha/\alpha$
within $z<0.14$ and $z<0.45$ is indicated with dotted lines.}
\label{fig3}
\end{center}
\end{figure}

We now turn to our second example, shown in Fig.~\ref{fig3}, in which $\mm=10.5,$ $\lambda=6,$
$B=45.4,$ and
$\kappa = 2.1 \times 10^{-4}.$ This solution passes muster on several counts: (1) The fit to the QSO
data is acceptable -- $\chi^2/d.o.f = 9.8/7$, corresponding to 20\% CL. (2) As can be seen
in the embedded box in Fig.~\ref{fig3}, the low-$z$ model results are in good agreement with both Oklo and
meteorite constraints. It is also consistent with equivalence principle bounds, the value of $\kappa$
being more than a factor of three below the limits discussed above. It should be stressed that
{\it less than
an order of magnitude improvement in the experimental sensitivity for composition-dependent inertial forces
may provide a direct test for this type of model.} As in the previous case, this solution is also consistent
with limits imposed by atomic clocks: we find $\phi^{\prime}_0 = -2.4 \times 10^{-3},$ giving
$\dot\alpha/\alpha = 3.5\times 10^{-17}$ yr$^{-1}$. (3) Finally, the model also gives variations
in $\alpha$ during the BBN and recombination epochs which are well within present sensitivities:
for $z=1100$ (CMB),
we find (retaining the linear expansion of $Z_F(\phi)$) that
$\Delta\alpha/\alpha = -4.4\times 10^{-4}$; for $z=10^{10}$ (BBN), we obtain
$\Delta\alpha/\alpha = -0.0025.$

\section{Discussion and Conclusions}

\noindent (1)~~In this work we have attempted to associate the possible temporal variation of $\alpha,$
the electromagnetic
fine structure constant, as indicated in absorption spectra of QSOs,
with the temporal evolution of the quintessence field which is responsible
for the present dark energy content of the visible universe. The outstanding obstacle in
maintaining this association is the observation that $\Delta \alpha,$ the deviation of   $\alpha$ from
its present value, is much smaller  in  the recent past (say $z<0.45$) compared to the
variation indicated by the higher-$z$ QSO data. The reconciliation can be effected
if the quintessence field has undergone a rapid slowing in the recent past. The quintessence
model that we study (the Albrecht-Skordis model) has precisely such a property.
We illustrate our results with an example (Fig.~\ref{fig3}) which can simultaneously fit the QSO
data and comply with the upper bounds on $\Delta \alpha$ from the Oklo and meteorite analyses
and atomic clock measurements. This class of solutions presents several inevitable predictions:
$(a)$ deviations from universal free fall should be observed when experimental sensitivity
is improved by a factor or 10 $(b)$ QSO measurements at ever-larger redshifts should
continue to show a monotonic decrease in $\Delta\alpha:$ in this type of solution
the quintessence field is just receding from its first turning point in its damped oscillation about
the fixed point. Another type of behavior (example in Fig.~\ref{fig1}) is seen in solutions which do
not satisfy the meteorite bound: there the quintessence field has completed a few oscillations,
allowing a return of $\alpha$ to its present value at large redshift.

\noindent (2)~~The evolution of the quintessence field, and therefore of $\alpha,$ is determined
by the quintessence potential. As noted in the text, this is in contrast to Bekenstein-type models
where changes in $\alpha$ are driven by the matter density. The latter allows local spatial
variation~\cite{Barrow:2002zh}, which could provide different dynamics for $\Delta\alpha$
in our local environment (Oklo and meteorites) and over cosmological scales (QSO data).
In the quintessence model, only the zero mode is relevant, and this option of spatial variation
is much suppressed.

\noindent (3)~~Comparison with Ref.~\cite{Gardner:2003nw}, in which the evolution of $\phi$ is driven
by both matter and by a harmonic scalar potential, shows that we require a much larger value
of $\mm$ for agreement with data. The origin of this difference is of interest, since
it highlights the constraints imposed by requiring that $\phi$ fulfill its role as a quintessence
field. Small values of the mass $(\alt H)$ will flatten the potential well to such an extent
that the field escapes entrapment and leads to an exit from the de Sitter phase. Moreover,
in our model, $\phi$ is not initially placed at the equilibrium point, and thus for a shallow well
will have considerable velocity during the present era, leading to a
strong disagreement with the low $z$ data.

\noindent (4)~~Because  the coupling $\kappa$ of the scalar field to
the electromagnetic Langrangian plays a dual role in determining both the variation of $\alpha$
and the violation of the equivalence principle, planned experiments on universal free fall
can directly test the viability of our model. These include the mission MICROSCOPE from the
Centre National d'Etudes Spatiales (CNES) expected to fly in 2005~\cite{MICROSCOPE}, and the National
Aeronautics and Space Agency (NASA) and European Space Agency (ESA) mission STEP (Satellite Test of
the Equivalence Principle)~\cite{STEP}.

\noindent (5)~~Our prediction for the present rate of variation of $\alpha$ is an order
of magnitude below present atomic clock sensitivities. Improvements in the accuracy
of such measurements to the range of $10^{-16}$ yr$^{-1},$ of the order of our result,
are anticipated for the near
future~\cite{Marion:2002iw}.

\noindent (6)~~Amusingly, we can project the evolution of $\alpha$  forward in time.
For our model in Fig.~\ref{fig3}, $\phi$ is at a turning point  in the
potential well of $V$ and consequently $\phi$ (and $\alpha$) would reverse motion in the future.
For the model in Fig.~\ref{fig1}, $\phi$ is already decreasing at present and will continue to
do so in the near future.

\hfill

\section*{Acknowledgments}
As always discussions with Carlos Nu\~nez were very valuable and enjoyable.
The work of LAA and HG has been partially supported by the US
National Science Foundation (NSF), under grants No.\ PHY--0140407
and No.\ PHY--0073034, respectively.



\begin{thebibliography}{99}
\bibitem{alpha} Throughout the paper we will omit any subscript on
the fine structure constant,
with the understanding that $\alpha$ denotes the present value.
\end{thebibliography}
\end{document}